\documentclass{appolb}
\input epsf.tex

\def\ni{\noindent}

\newcount\rys

\def\rysunek#1#2{
\global\advance\rys by 1 \vbox{\centerline{\epsfxsize=4.5truein\epsfbox{#1}}
\vskip 0.1truein
\centerline{\vtop{\ni Fig. \the\rys . #2}}}
\vskip 0.2truein minus 0.2truein}

\def\podwrys#1#2#3{
\global\advance\rys by 1 \vbox{
\centerline{\epsfxsize=2.5truein\epsfbox{#1}}
\vskip 0.1truein
\centerline{\epsfxsize=2.5truein\epsfbox{#2}}
\vskip 0.1truein
\centerline{\vtop{\ni Fig. \the\rys . #3}}
\vskip 0.1truein}
}

\def\poczwrys#1#2#3#4#5{\vskip 0.2in minus 0.2in
\global\advance\rys by 1 \vbox{
\centerline{\epsfxsize=2truein\epsfbox{#1} \hskip 0.1truein
\epsfxsize=2truein\epsfbox{#2}}
\vskip 0.1truein
\centerline{\epsfxsize=2truein\epsfbox{#3} \hskip 0.1truein
\epsfxsize=2truein\epsfbox{#4}}
\vskip 0.1truein
\centerline{\vtop{\ni Fig. \the\rys . #5}}}
\vskip 0.2in minus 0.2in}

\def\poczwrysl#1#2#3#4#5{\vskip 0.2in minus 0.2in
\global\advance\rys by 1 \vbox{
\centerline{\null\hskip -0.2truein\epsfxsize=2truein\epsfbox{#1} \hskip 0.5truein
\epsfxsize=2truein\epsfbox{#2}}
\vskip 0.1truein
\centerline{\null\hskip -0.2truein\epsfxsize=2truein\epsfbox{#3} \hskip 0.5truein
\epsfxsize=2truein\epsfbox{#4}}
\vskip 0.1truein
\centerline{\vtop{\ni Fig. \the\rys . #5}}}
\vskip 0.2in minus 0.2in}

\title{
\centerline{\bf THE FIRST COMPACT OBJECTS}
\vskip 0.2in
\centerline{\bf IN THE MOND MODEL\footnote{
This research is partially supported by the Polish State Committee
for Scientific Research (KBN), grant no. 2 P03B 112 17.
}}
}
\author{
S. Stachniewicz$^a$
\and
M.  Kutschera$^{a,b}$
\address{
$^a$H.Niewodnicza\'nski Institute of Nuclear Physics, ul. Radzikowskiego 152,\\
31-342 Krak\'ow, Poland\\
$^b$Institute of Physics, Jagiellonian University, ul. Reymonta 4,\\
30-059 Krak\'ow, Poland}
}

\begin{document}
\maketitle

\begin{abstract}
We trace the evolution of a spherically symmetric density perturbation
in the MOdified Newtonian Dynamics (MOND) model. The background cosmological
model is a $\Lambda$-dominated, low-$\Omega_b$ Friedmann model with no Cold
Dark Matter. We include thermal processes and non-equilibrium chemical
evolution of the collapsing gas. We find that the first density
perturbations which collapse to form luminous objects have mass
$\sim 10^5 M_{\odot}$. The time of the final collapse of these objects
depends mainly
on the value of the MOND acceleration $a_0$ and also on the baryon density
$\Omega_b$. For the "standard" value $a_0=1.2\times 10^{-8}$ cm/s$^2$ the
collapse starts at redshift $z \sim 160$ for $\Omega_b=0.05$ and $z \sim 110$
for $\Omega_b=0.02$.
\end{abstract}

\PACS{95.30.Lz, 95.30.Sf, 98.35.Mp, 98.80.Bp}

\section{Introduction}

Recent developments in cosmological observations have led to so-called
cosmological concordance model with $\Omega_b$ about 0.03, $\Omega_m$
(dark+baryonic) about 0.3 and $\Omega_{\Lambda}$ about 0.7. However,
as the $\Lambda$CDM models are dominated by hypothetical vacuum energy
and non-baryonic dark matter contributions, some scientists look for
different solutions.
Perhaps the most intereresting alternative model is the MOdified
Newtonian Dynamics model (MOND) proposed by M.Milgrom\cite{Mil83}.
It assumes that there is no non-baryonic dark matter (or it is negligible)
and the lack of matter is only apparent due to modification of
dynamics or gravity for small accelerations ($a\ll a_0$ where $a_0$ is
some constant). This model seems to work very well for spiral galaxies
and many other types of objects\cite{Mil99} but, however, it has some
unresolved problems (e.g. lack of covariance).

Our aim there is to study what would be the implications of the MOND model
for the formation of the very first objects in the Universe.

\section{MOND vs the standard theory of linear perturbations}

To apply the MOND model to structure formation calculations one
encounters a number of difficulties. First of all, MOND is not a
theory, it is rather a phenomenological model. In its present form
MOND is inconsistent with the General Relativity. Up to now there were
no successful attempts to find a generally covariant theory that could
be a generalisation of the General Relativity and would give a MOND-like
predictions in the low-gravity limit \cite{Mil99}.

MOND is a model that modifies either dynamics or gravity (in this paper we
assume this second possibility). It introduces a new fundamental scale,
usually
called $a_0$. Gravitational fields much stronger than $a_0$ are identical to
their Newtonian limit $g_N$ and very weak fields are $\sqrt{a_0 g_N}$.
According to Sanders and Verheijen \cite{SVe98} the value of the fundamental
acceleration scale is $a_0 = 1.2 \times 10^{-8} \rm{cm}/\rm{s}^2$. More
precisely, the strength of the gravitational field may be written as

\begin{equation}
\mu\left({g \over a_0}\right) \vec{g} = \vec{g}_N ,
\end{equation}

\ni where $\mu(x)$ is some function that interpolates between these
two extreme cases. This function is not specified in the model.
We have decided to apply the function used by Sanders and Verheijen
\cite{SVe98}:

\begin{equation}
\mu(x) = {x \over \sqrt{1+x^2}}
\end{equation}

\ni and, finally,

\begin{equation}
\vec{g} = \vec{g}_N \sqrt{1+\sqrt{1+({2 \over x})^2} \over 2}
\end{equation}

\ni where $x = g_N/a_0$.

If we consider the gravitational field of a point mass $M$, for distances
$R>\sqrt{GM/a_0}$ the gravitational field would be in the MOND regime,
where its strength would decrease as $1/R$ instead of $1/R^2$. It means
that there is no escape velocity and all systems are gravitationally bound.

The conseqences of the MOND for cosmology are not studied in details yet.
R.H.Sanders \cite{San98} suggested that because in the early Universe
the MOND radius is much lower than the radius of the horizon the evolution
of the scale factor is described by the standard Friedmann equations.
Here we follow this assumption and study the formation of the first
objects in the Universe with modified dynamics.

\subsection{Collapse of a pressureless fluid in MOND.}

Let us consider a homogenous ball of density $\varrho$ and some radius $R$,
expanding uniformly in all directions with speed proportional to the
distance from the center. For a sphere of radius $r$ the deceleration in the
Newtonian limit is

\begin{equation}
g_N = {GM \over r^2} = {4 \over 3} \pi G \varrho r
\end{equation}

\ni and, of course, expansion is scale-invariant because deceleration
and velocity always are proportional to the radius $r$. However,
we can find some radius $r_0$ where $g_N(r)<a_0$ for $r<r_0$ -- further we
will call that a `MOND radius'. For $r<r_0$ the gravity is in the MOND
regime and the dynamics is changed. The evolution of the MOND radius $r_0$ in
the early Universe is discussed in details by Sanders \cite{San98}.

In the standard perturbation theory, if the mean density is comparable with
the critical density of the Universe (it is true at least for large redshifts)
the recollapse depends very strongly on the value of the overdensity.
Regions with mean density less than the critical density will not
recollapse at all and vice versa. Moreover, the time of the recollapse
is very sensitive to the value of the density. In MOND it is
different because there is no critical density. The overdensity does not
play an essential role and the recollapse is similar for regions of
different densities.

It is quite easy to derive the linear perturbation theory in the Newtonian
limit (e.g. in Kolb and Turner \cite{KTu90}). In MOND it is much more
difficult because there appear
nonlinear terms connected with $\nabla\varphi$, where $\varphi$ is the
gravitational potential. However, it is known that the rare highest
fluctuations in the primordial density field were nearly spherically
symmetric \cite{Eis95}, so as long as we are concerned with the very first 
bound objects in the Universe we may assume spherical symmetry.

Now let us consider a spherically symmetric overdensity with
density profile

\begin{equation}
\varrho(r) = (1+\delta(r))\bar{\varrho} .
\end{equation}

\ni Of course, the density changes with time because of expansion or
recollapse, but the mass inside some shell $i$

\begin{equation}
M_i = \int\limits_0^{r_i} 4\pi r^2 \varrho(r) dr
\end{equation}

\ni remains constant, so the deceleration for a shell of radius $r_i$
will be equal to

\begin{equation}
{d^2 \over dt^2} r_i = -f(r_i)
\end{equation}

\ni where $f(r_i)$ is the MOND gravitational force and it depends on
$GM_i/r_i$ and $a_0$. Let us drop the subscript $i$. If we multiply
this equation by $dr/dt$ we get

\begin{equation}
{d \over dt} \left[ {1\over 2} \left( {dr\over dt} \right)^2 \right] =
-f(r) {dr\over dt}
\end{equation}

\ni and after integrating over $t$ we obtain

\begin{equation}
{dr\over dt} = \sqrt{-2F(r) + C} ,
\end{equation}

\ni where $F'(r)=f(r)$ and $C$ is some constant which may be easily
calculated if we know initial radius and velocity for a given function
$F(r)$. This formula may be easily integrated, e.g. with the Runge-Kutta
algorithm.

To show the difference between the MOND and the Newtonian gravity,
we have performed a set of calculations. We have traced the evolution
of a dust cloud (no pressure) expanding homogenously in all directions.
Initial radius $r_i$, velocity $v_i$, density $\varrho_i$ and time $t_i$
were taken from the $\Omega=1$ Friedmann model with no radiation for
$z+1=500$ and $h=0.65$. Initial radius was chosen to be the MOND radius at
that time. We have performed eight runs:

\begin{itemize}

\item pure Newtonian gravity with the density equal to $\varrho_i$,
$0.80\varrho_i$ and $1.25\varrho_i$

\item MOND with the density as above

\item MOND with the $a_0$ parameter ten times greater than the standard
value and mean density equal to $\varrho_i$

\item MOND with the $a_0$ parameter ten times lower than the standard
value and mean density equal to $\varrho_i$

\end{itemize}

The results are displayed at Fig. 1. Solid curves show the
trajectories in the MOND model and the long-, middle- and
short-dashed ones show the trajectories for the Newtonian gravity, MOND with
large $a_0$ and MOND with low $a_0$, respectively. For the ``standard'' MOND
and the Newtonian gravity lower curves show the trajectories for runs with
greater densities ($1.25\varrho_i$) and upper curves show the trajectories
for lower densities (0.80$\varrho_i$).

\rysunek{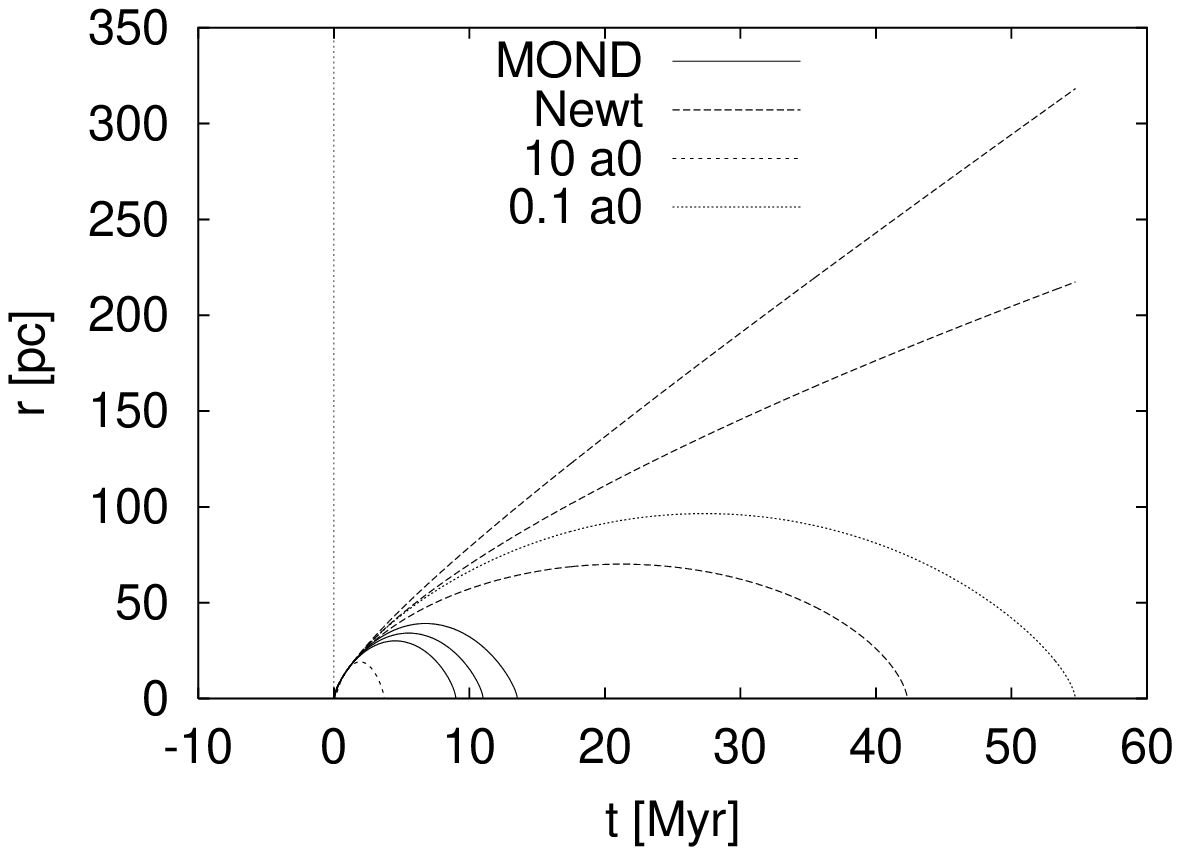}{Collapse of a pressureless fluid in
MOND and in the Newtonian gravity}

As one could expect, raising or lowering the initial density by 25\%
does not have a big influence if we apply the MOND model while it plays a
crucial role for the Newtonian gravity if the density is near the critical
one (it is always true in the early Universe which is flat at least
asymptotically at the beginning) -- models with higher densities tend to
recollapse while models with the lower one do not. The results are quite
sensitive to the value of $a_0$ because it is the limit between the Newtonian
and the MOND regimes -- however, it is a quantitative effect only and the
dust clouds in such models will always recollapse.

\subsection{Collapse of a perfect gas.}

If we take a perfect gas instead of the pressureless fluid, the evolution
will look different because effects of pressure will moderate the recollapse,
especially for small systems. As we assume spherical symmetry, we use
Lagrangian coordinates.

The dynamics is governed by the following equations:

\begin{eqnarray}
{dM \over dr} & = & 4\pi r^2 \varrho, \label{ciaglosc}\\
{dr \over dt} & = & v , \label{promien}\\
{dv \over dt} & = & -4\pi r^2 {dp \over dM}-{GM(r) \over r^2} ,
\label{predkosc}\\
{du \over dt} & = & {p \over \varrho^2} {d \varrho \over dt} + {\Lambda \over 
\varrho} ,
\label{energia}
\end{eqnarray}

\ni where $r$ is the radius of a sphere of mass $M$, $u$ is the internal
energy per unit mass, $p$ is the pressure and $\varrho$ is the mass
density. Here eq.(\ref{ciaglosc}) is the continuity equation, (\ref{promien})
and (\ref{predkosc}) give the acceleration and (\ref{energia}) accounts for
the energy conservation.
The last term in the eq.(\ref{energia}) describes cooling/heating
of the gas, with $\Lambda$ being the energy absorption (emission) rate per
unit volume, given in details in \cite{Sta01}.

We use the equation of state of the perfect gas

\begin{equation} p= (\gamma -1) \varrho u , \end{equation}

\ni where $\gamma = 5/3$, as the primordial baryonic matter after 
recombination
is assumed to be composed mainly of monoatomic hydrogen and helium, with
the fraction of molecular hydrogen $H_2$ always less than $10^{-3}$.

In case of modified gravity, equation (\ref{predkosc}) will look a bit
different:

\begin{equation}
{dv \over dt}= -4\pi r^2 {dp \over dM}-f\left({GM(r) \over r^2}\right) ,
\end{equation}

\ni where $f(x)$ is the function inverse to the function $\mu(x)$ mentioned
before and it is asymptotically equal to $x$ for $x\gg a_0$ and $\sqrt{a_0x}$
for $x\ll a_0$.

\section{Code used in the simulations}

In the simulations we have used the code described in \cite{Sta01}, based on
the codes described by Thoul and Weinberg \cite{Tho95} and Haiman, Thoul
and Loeb \cite{Hai96}. This is a standard, one-dimensional, second-order
accurate Lagrangian finite-difference scheme. The only changes were
modification of gravity and putting the dark matter fraction $\Omega_{dm}$
equal to zero. However, it was necessary to make significant changes
in initial conditions.

First of all, we start our calculations at the end of the
radiation-dominated era. For $\Omega_b=\Omega_m=0.02$, $z_{eq}=203$ and for
$\Omega_b=0.05$, $z_{eq}=507$ as given by the formula provided by Hu and
Eiseinstein \cite{HE98},
$z_{eq}=2.50 \times 10^4 \Omega_0 h^2 \Theta_{2.7}^{-4}$,
where $\Theta_{2.7}=T_\gamma/2.7 K$, assuming $h=0.65$ and
$T_\gamma$=2.7277 K. We have assumed that, as in the standard
cosmology, initial overdensities may grow only in the matter-dominated
era. We have developed and tested our own
code to calculate the initial chemical composition and initial gas
temperature. We have compared our results with the results by Galli and Palla
\cite{Gal98} and they turned out to be very similar -- the agreement is at
a level of 10-20\%.
The difference was probably due to the fact that Galli and Palla
have included more species (e.g. deuterium and lithium) and some reaction
rates that they used were a bit different.

\section{Results}

We have performed eight runs, for various combinations of $\Omega_b$
(0.02 and 0.05), $\Omega_{\Lambda}$ ($1-\Omega_b$ and 0) and $a_0$
($1.2 \times 10^{-8} \mbox{cm/s$^2$}$ and $1.2 \times 10^{-9}
\mbox{cm/s$^2$}$). Total mass of a cloud is about
$3 \times 10^5$ M$_\odot$. Results are displayed on Figs. 2-7.

Fig. 2 shows trajectories of shells enclosing 7\%, 17\%, 27\% ... 97\%
of the total mass for the models with the ``standard'' value of $a_0$.

Fig. 3 shows the same for the models with $a_0$ 10 times lower than the
``standard'' value.

In figures 2-5 upper plots are for $\Omega_b=0.02$, lower ones for
$\Omega_b=0.05$, plots at left for $\Omega_{\Lambda}=1-\Omega_b$
and plots at right for $\Omega_{\Lambda}=0$.

\poczwrys{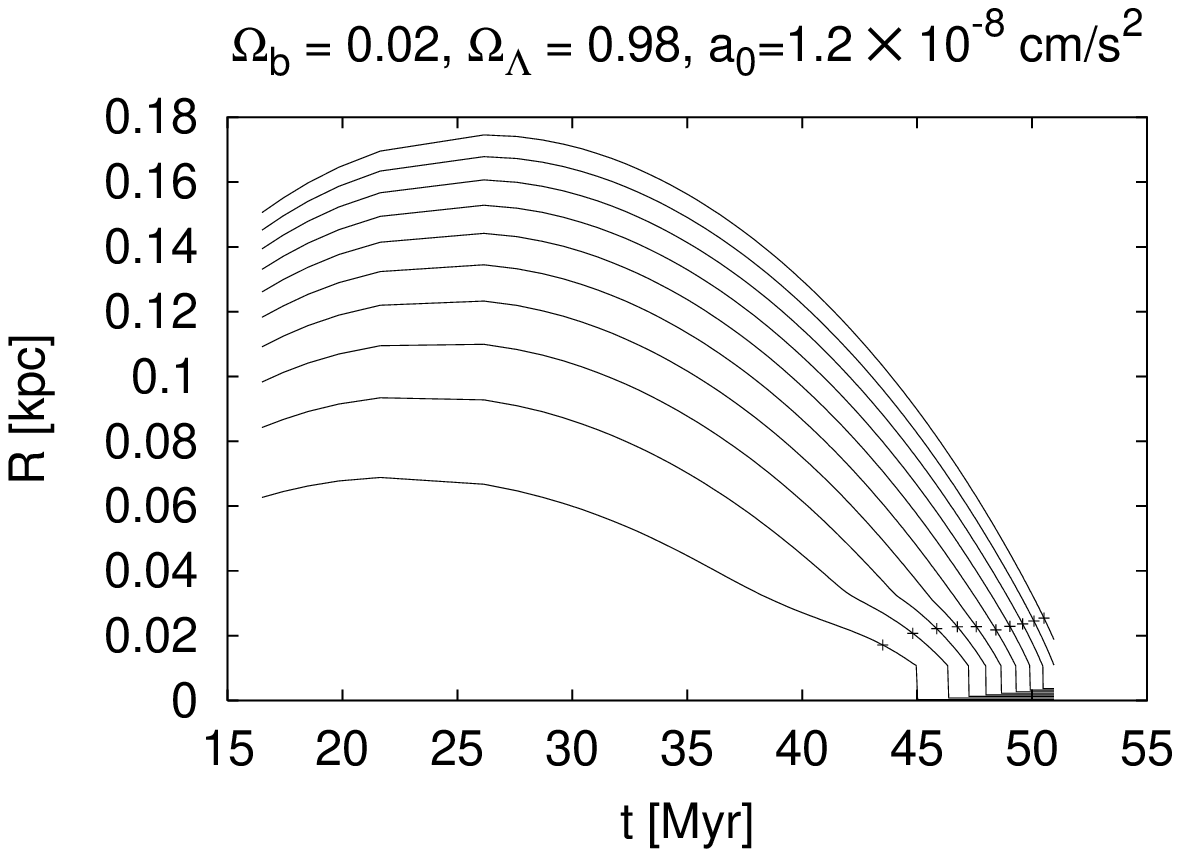}{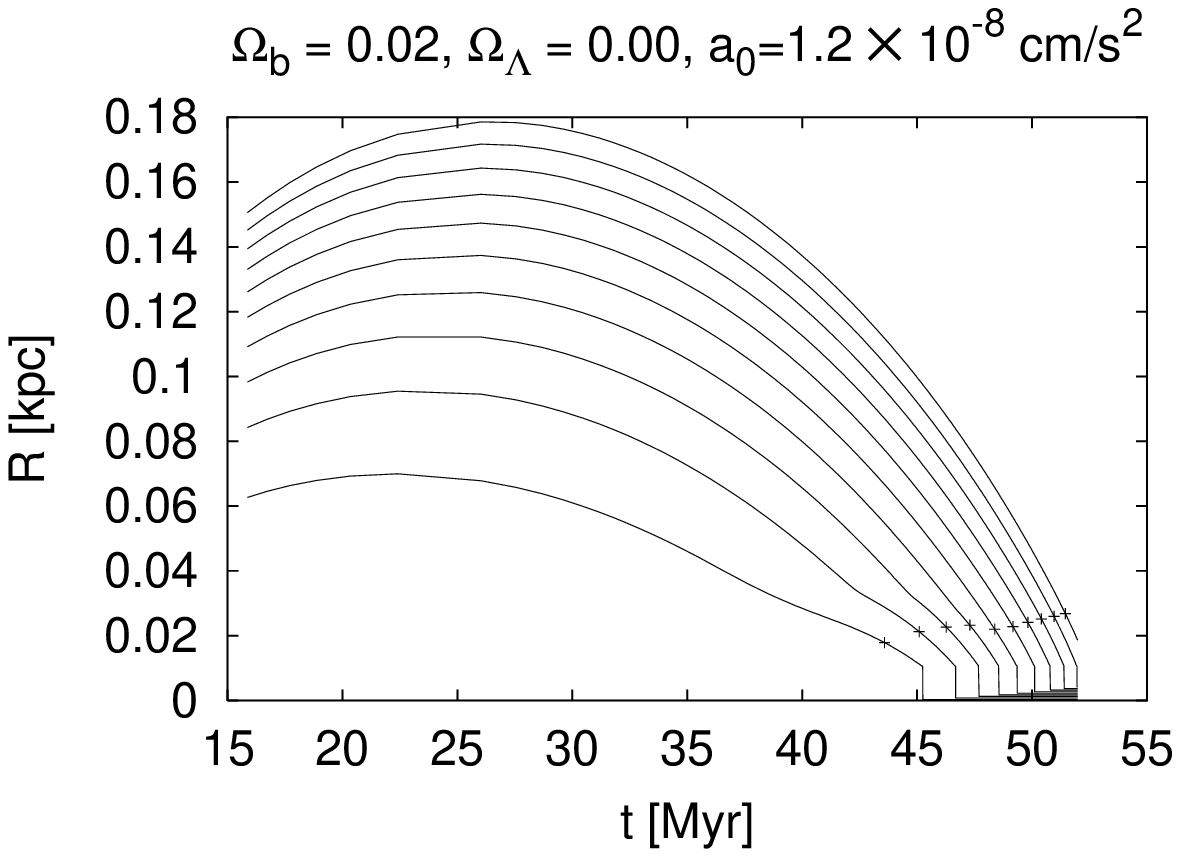}{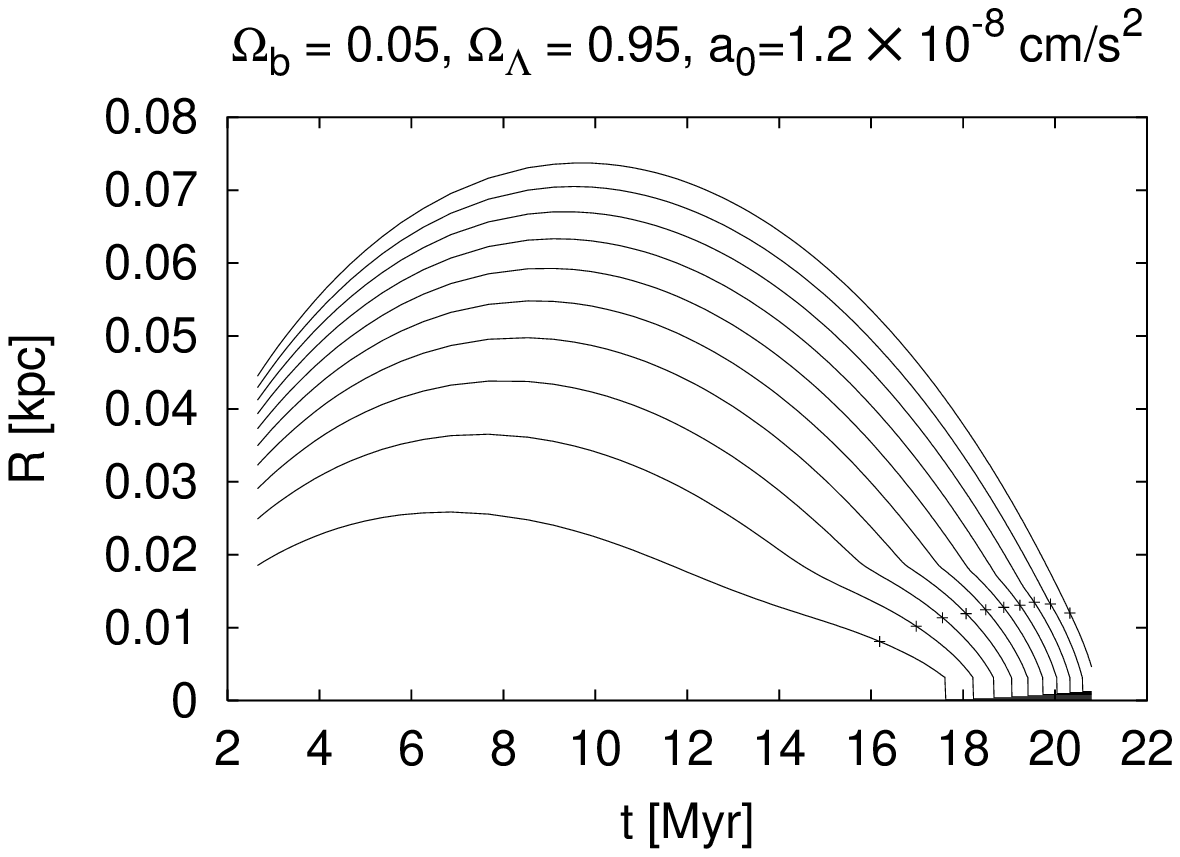}{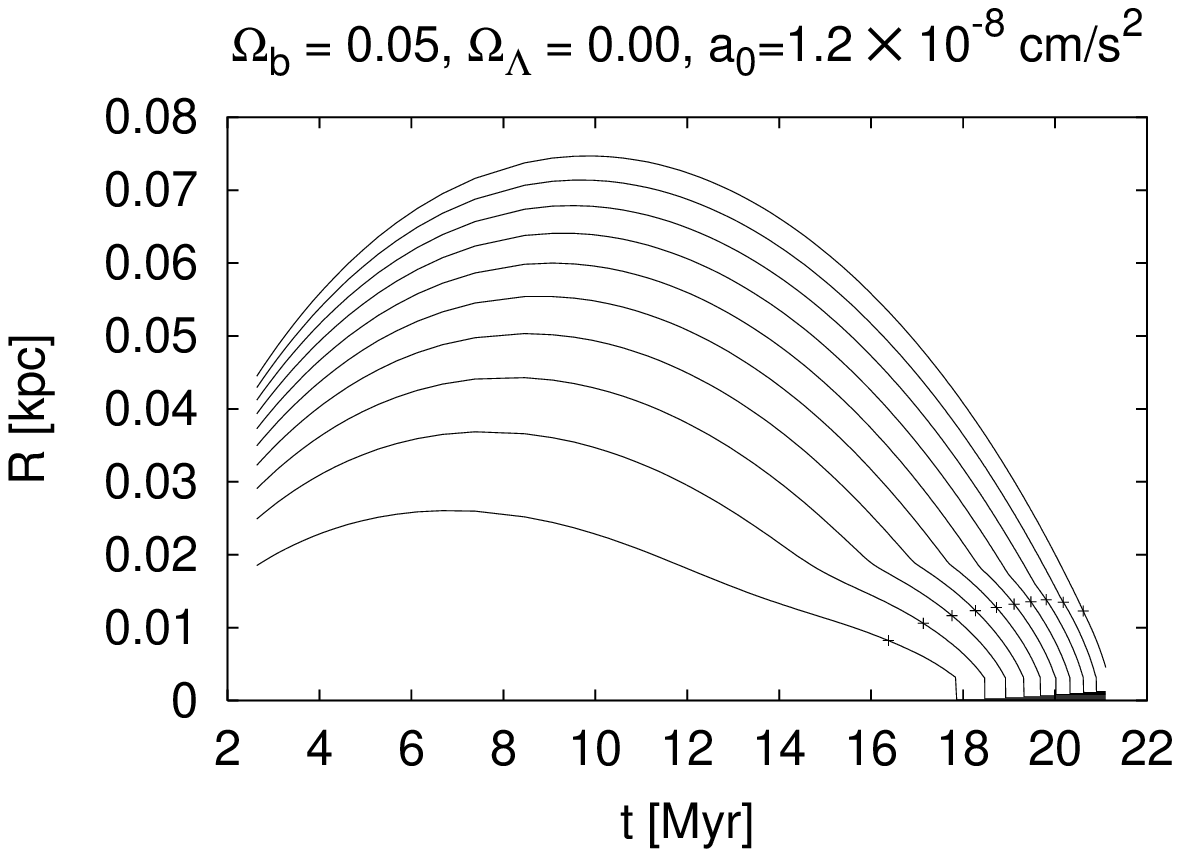}{Shell trajectories for
the first four runs, for the ``standard'' value of $a_0$.}

\eject
\poczwrys{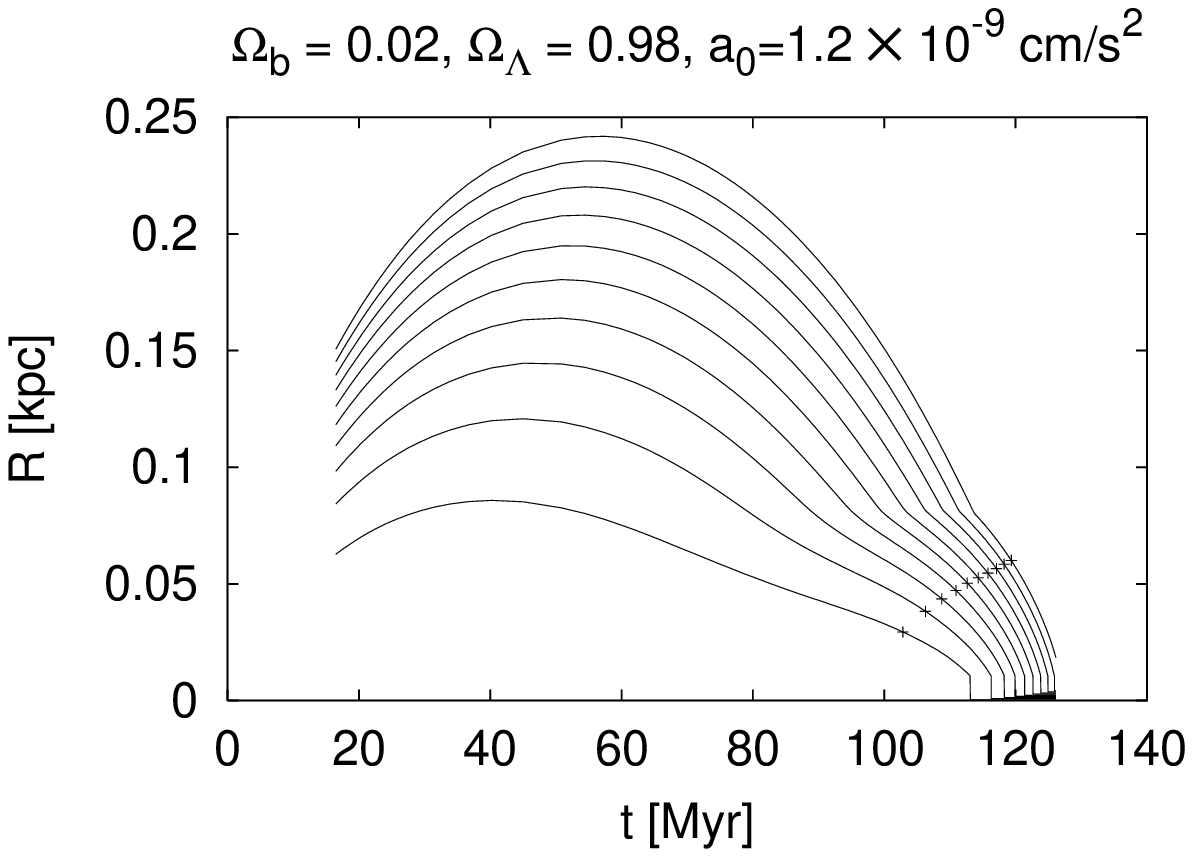}{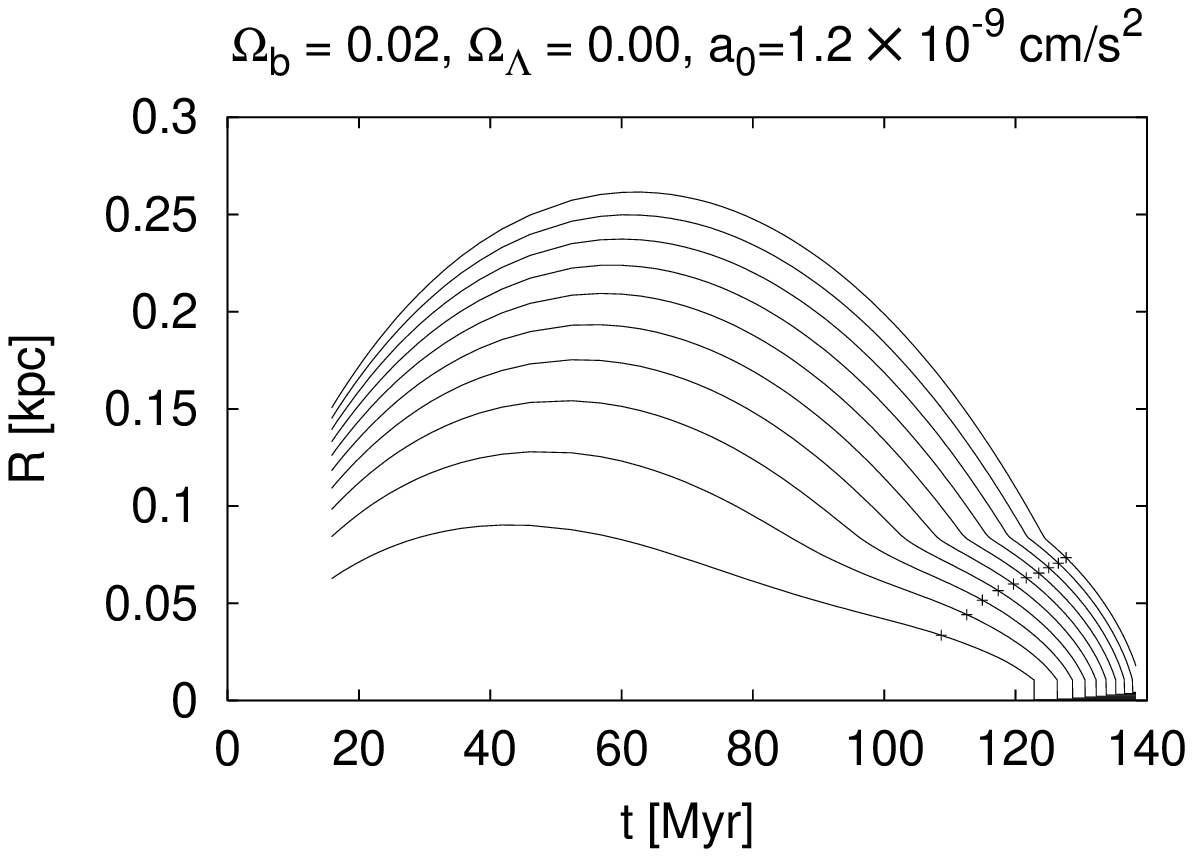}{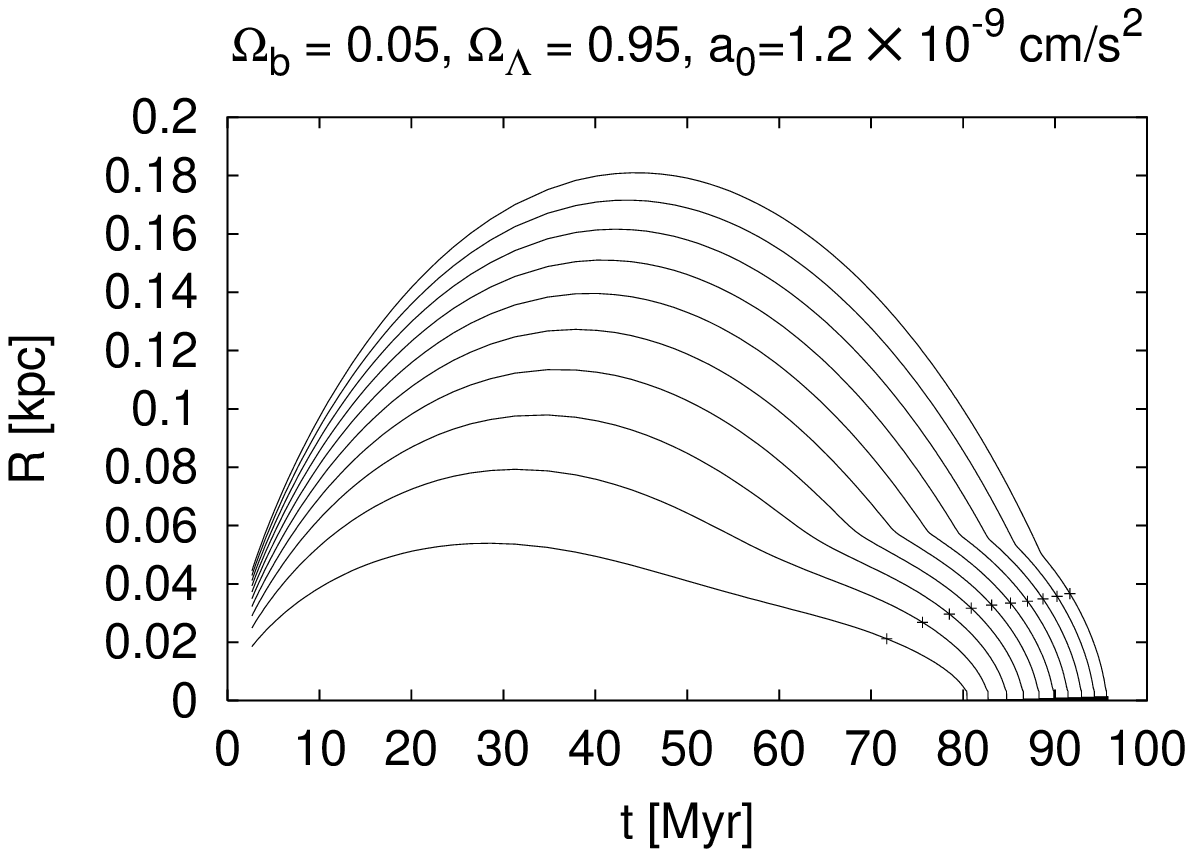}{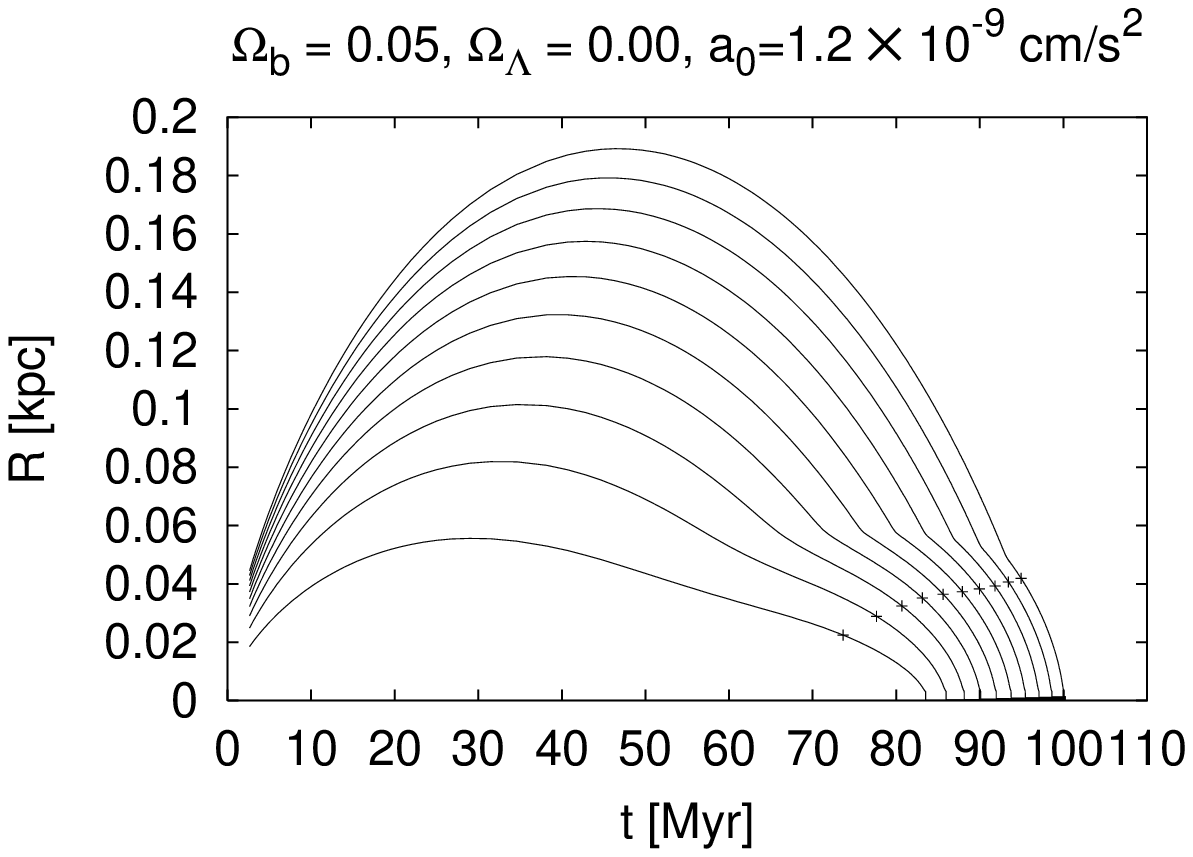}{Shell trajectories for
the last four runs, for $a_0$ 10 times lower than the ``standard'' value.}

These results show that:

\begin{itemize}

\item if we set the cosmological constant to zero, it affects the results
very slightly -- collapse is a bit slower (it is due to the fact that
non-zero cosmological constant affects behaviour of the scale factor and,
thus, the initial velocities)

\item time of collapse depends very strongly on $a_0$

\item results are more sensitive to the value of $\Omega_b$ for
higher $a_0$

\item mass $\sim 10^5$ M$_\odot$ is the boundary between the regime
when the gas effects dominate and the cloud virializes without further
collapse to a compact object (or at least this collapse is much slower), and
the collapse regime when the gas effects may be neglected. The boundary
mass is a bit higher for lower $a_0$.

\end{itemize}

The last point needs some comments. Scientists who explore the origin of
the Large Scale
Structure often neglect gas terms, i.e. set pressure and internal energy
equal to zero. Our paper is not the first one that deals with the structure
formation in MOND -- this problem was also explored by Sanders
(\cite{San98} and \cite{San00}) and  Nusser \cite{Nus01}
but they were interested mainly in the large structure formation and they did
not include gas effects. It is a good approximation for large scales but gas
effects play a crucial role in small scale structure formation.

If we look e.g. at Fig. 2, for the outer shells the collapse is almost
``symmetric'' to the expansion, the slow-down by gas pressure is very tiny.
In contrast, the most innermost shells show a big asymmetry -- expansion,
collapse and then shock waves make the matter to virialize (the kinetic energy
of a collapsing shell is turned into the internal energy of the gas) so the
collapse is stopped. Only pressure of the outer shells and the
cooling processes (especially the cooling by molecular hydrogen H$_2$)
force them to collapse.

Figs. 4 and 5 show the same as Figs. 2 and 3, but as a function of redshift
instead of the cosmic time. They show that the collapse of the first
compact objects in MOND is really very fast -- for high $\Omega_b$ and
``standard'' $a_0$ it happens at $z\sim 160$, for lower $\Omega_b$
it happens at $z\sim 110$ and for lower $a_0$ at $z\sim 50-60$.
It is a very strong prediction and hopefully it may be tested in the future
by the Next Generation Space Telescope (NGST) or (perhaps) some its
successors -- if they discover luminous objects
or ionized gas at $z>20-30$ it will strongly support the MOND model.

\poczwrys{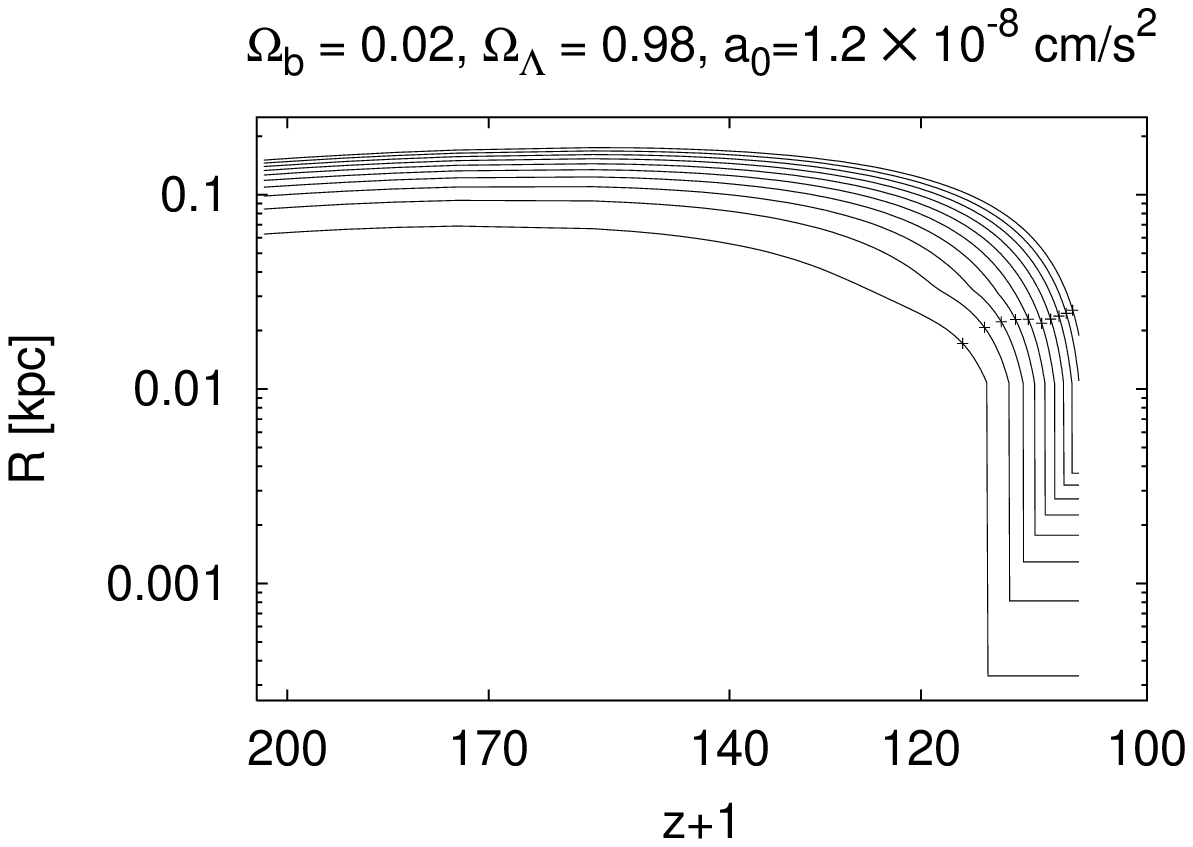}{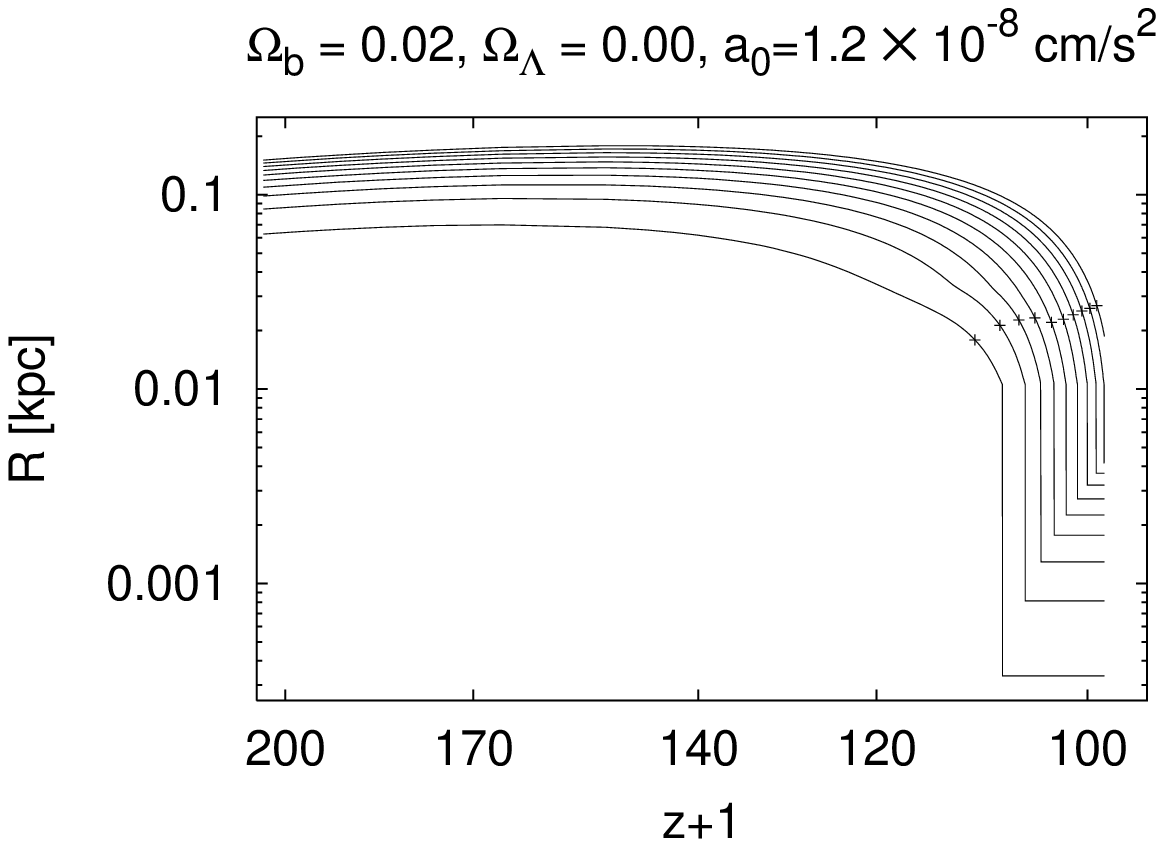}{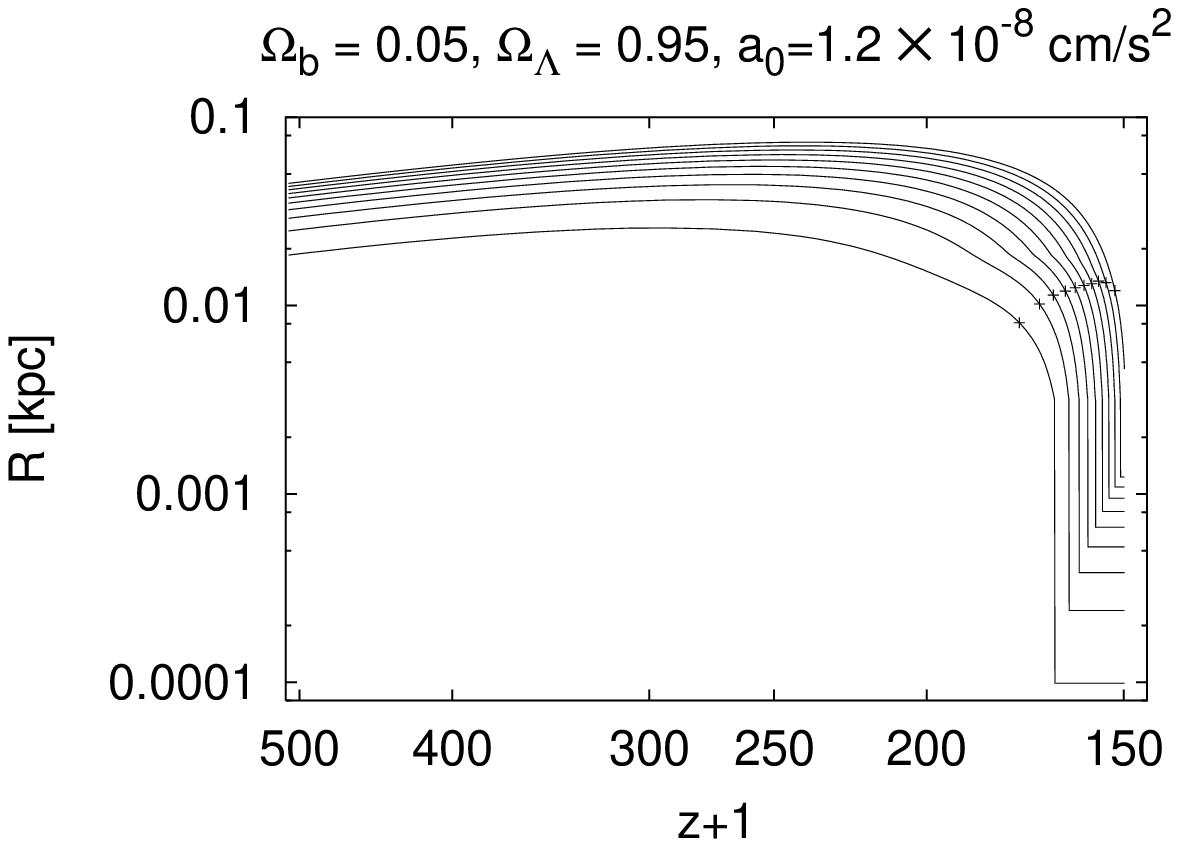}{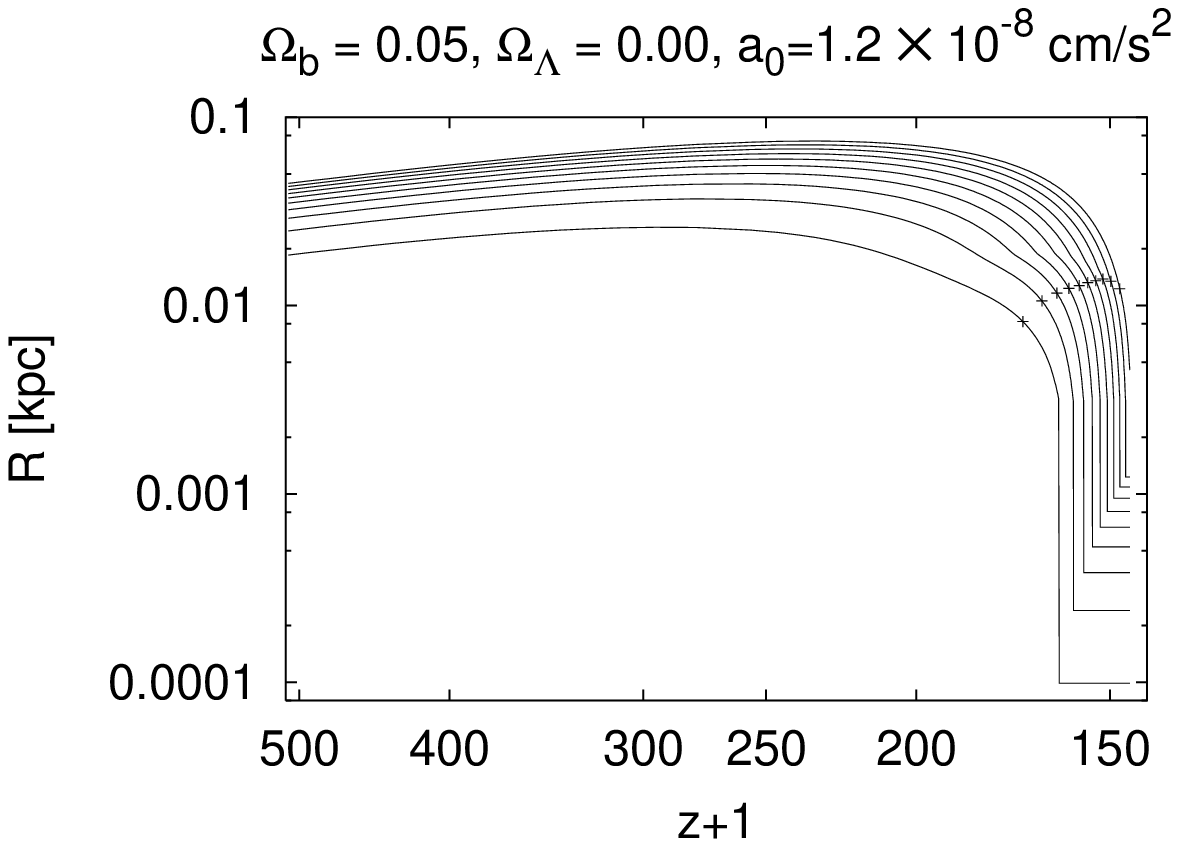}{Shell trajectories
as a function of redshift, for the first four runs or the ``standard'' value
of $a_0$.}

\poczwrys{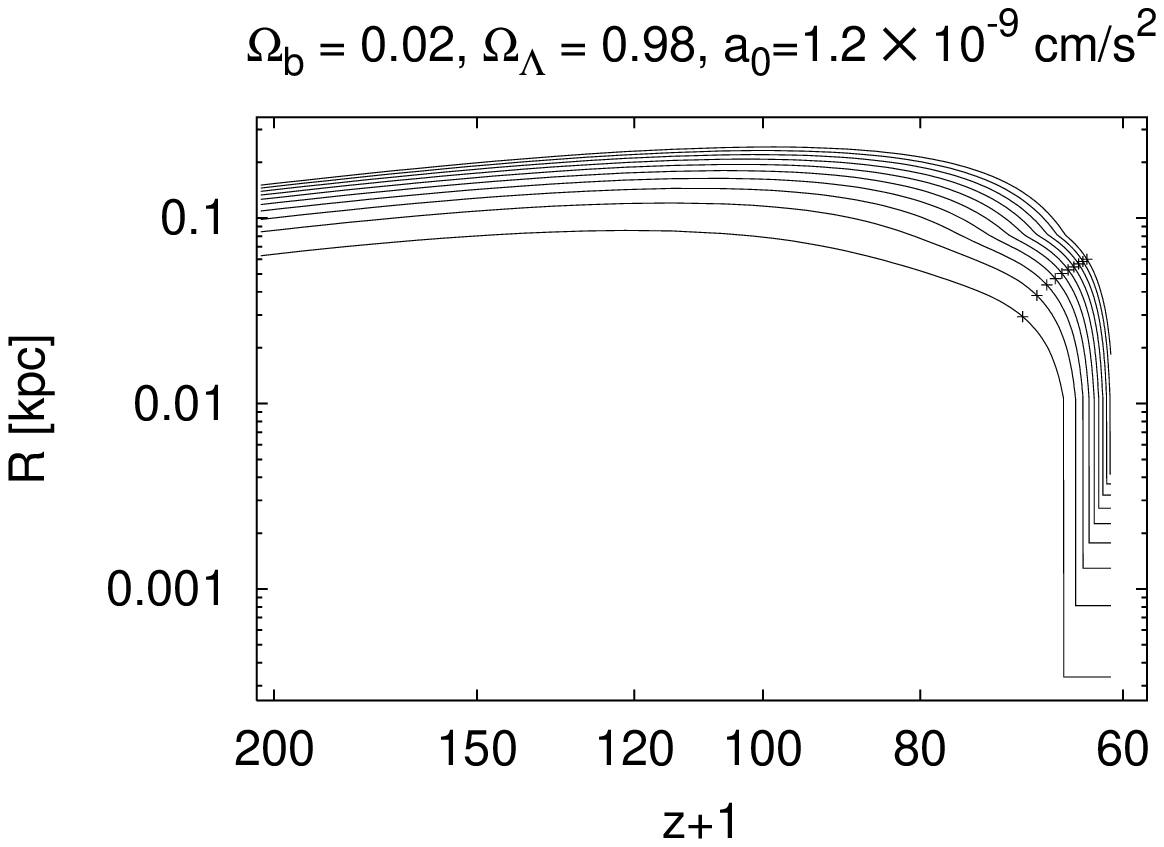}{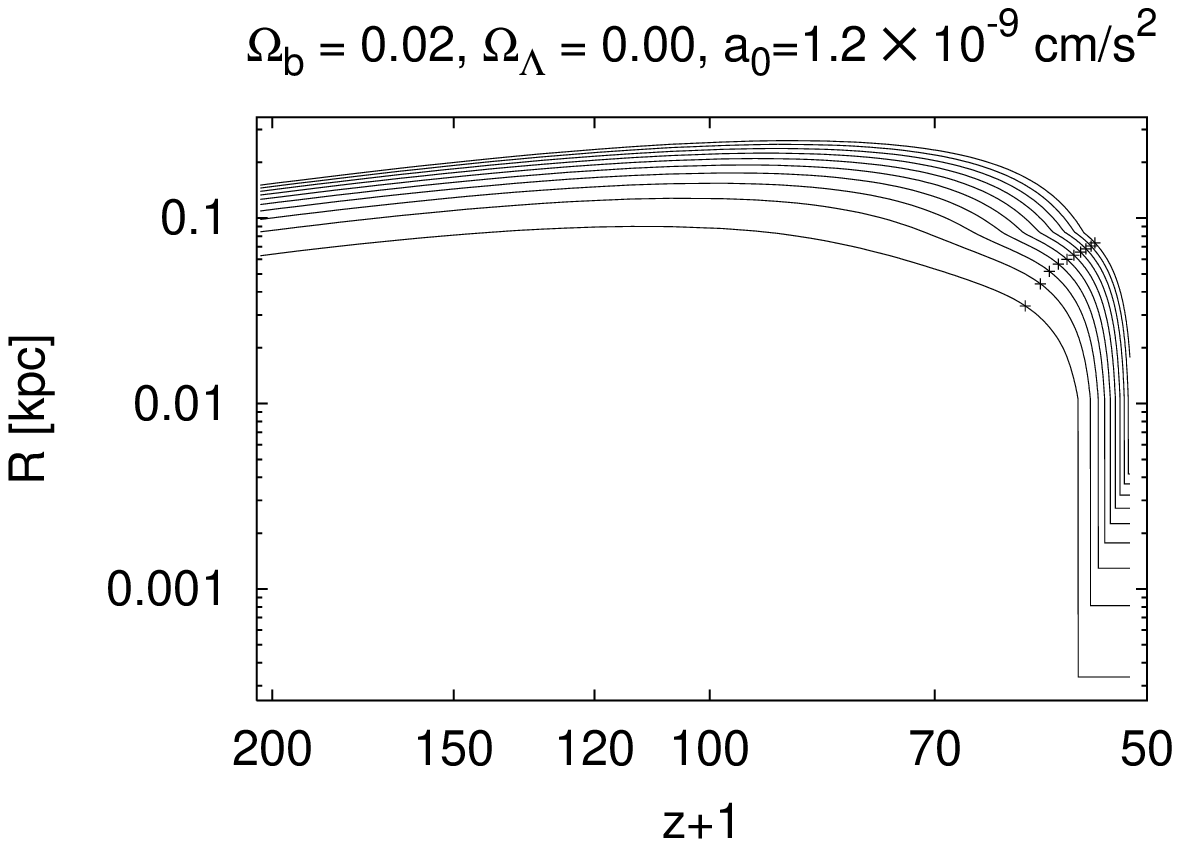}{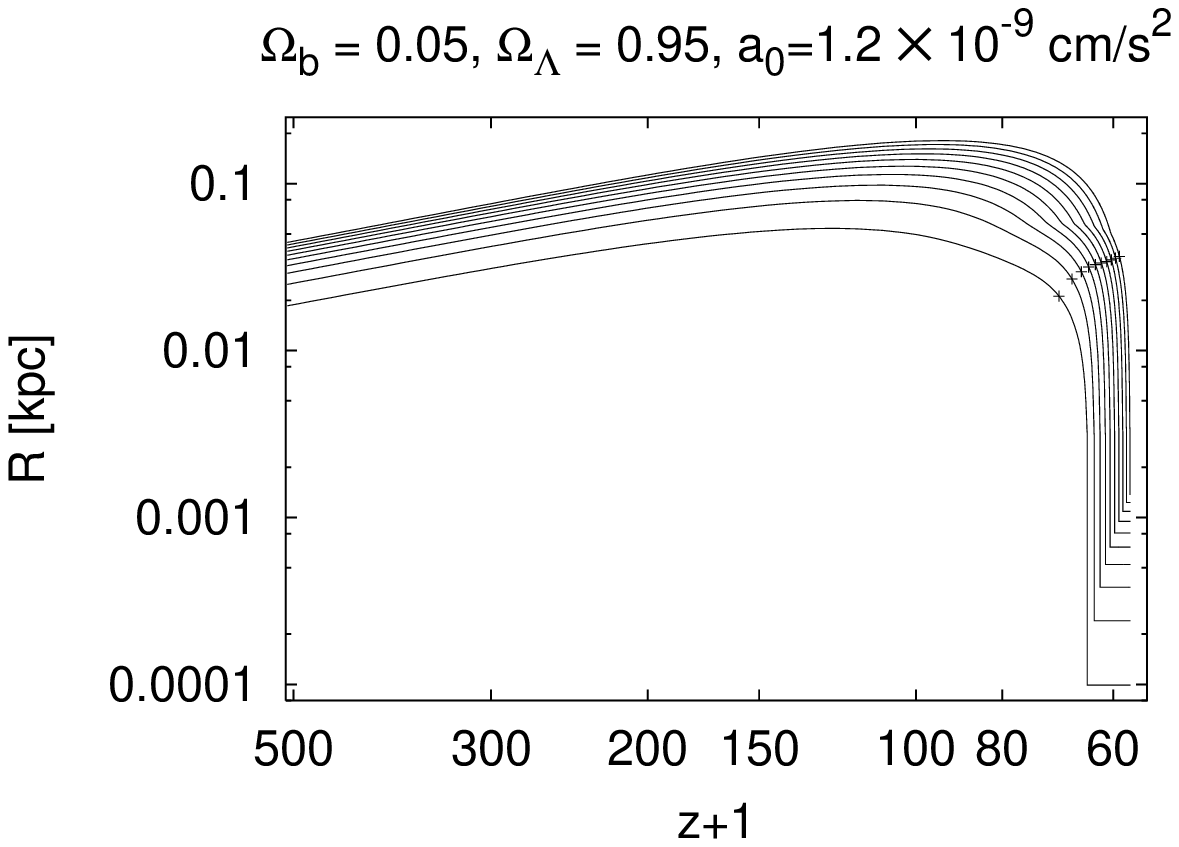}{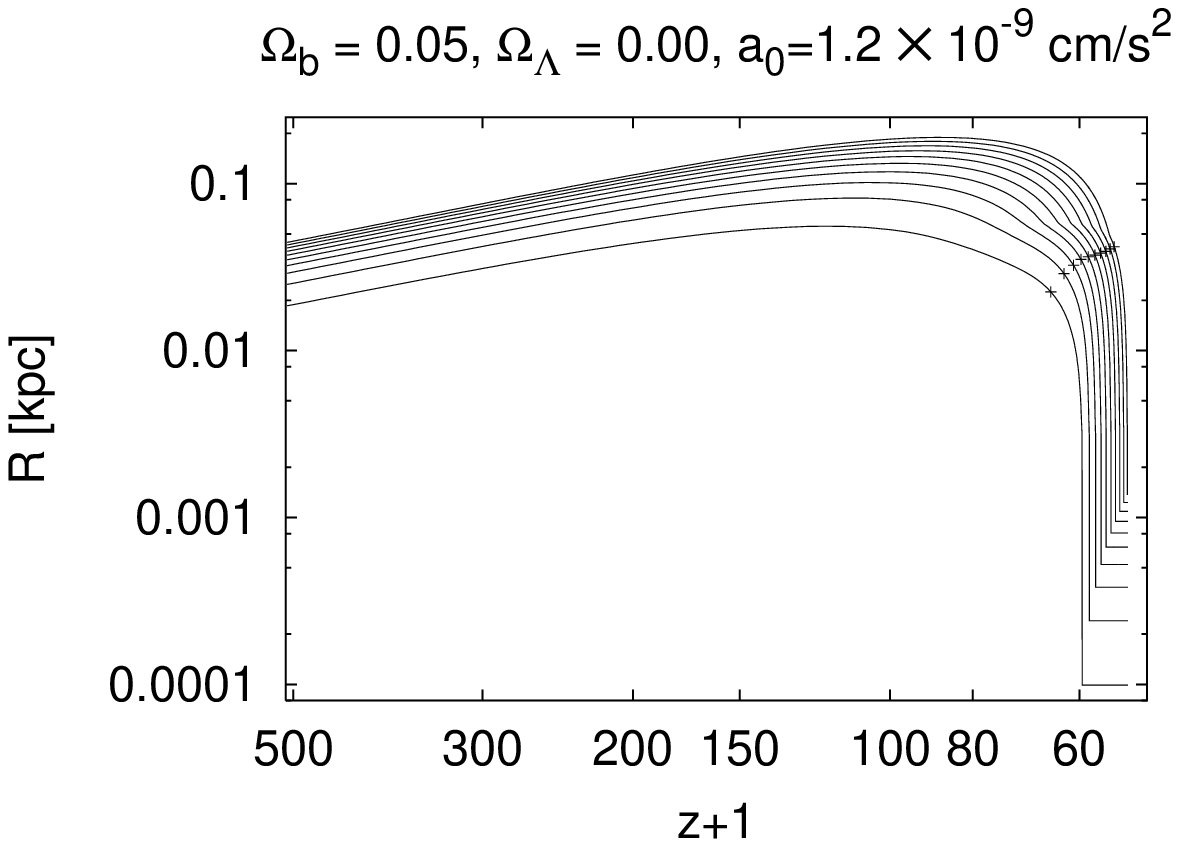}{Shell trajectories
as a function of redshift, for the last four runs or $a_0$ 10 times lower
than the ``standard'' value.}

\poczwrys{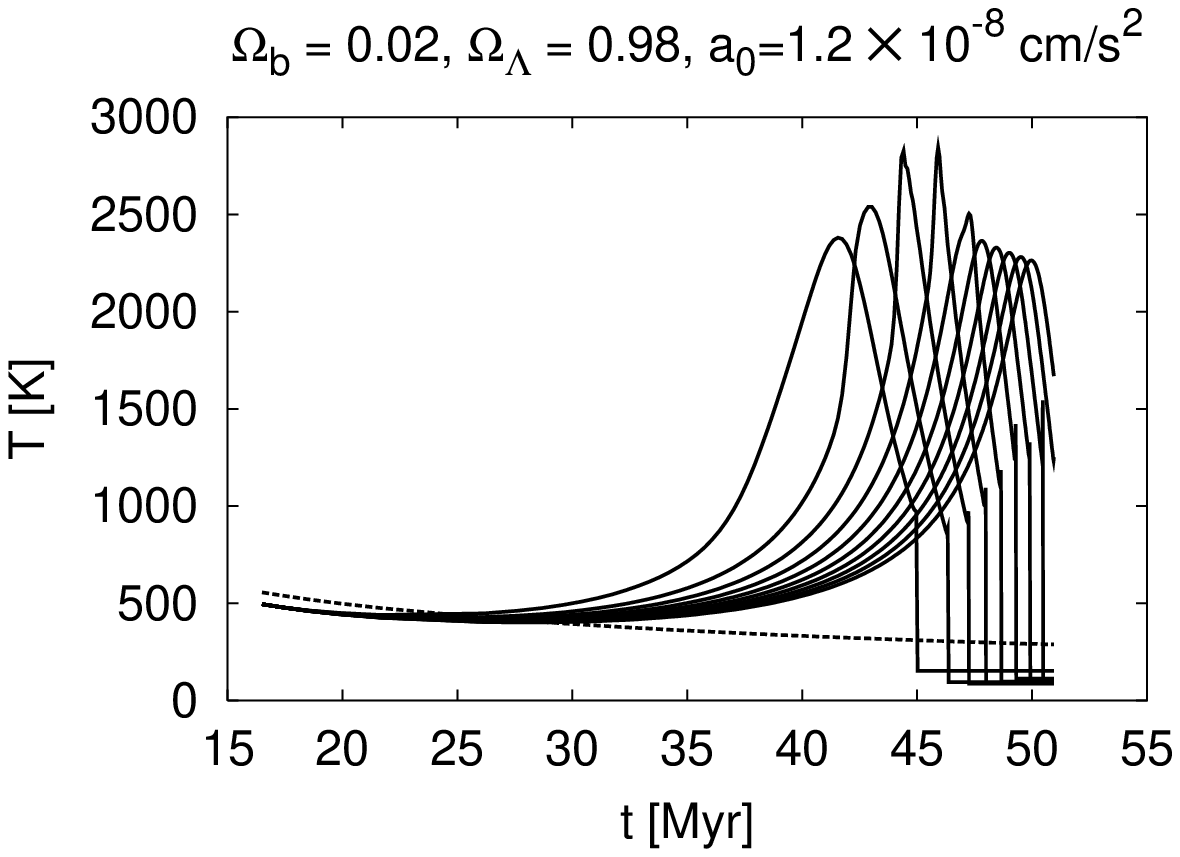}{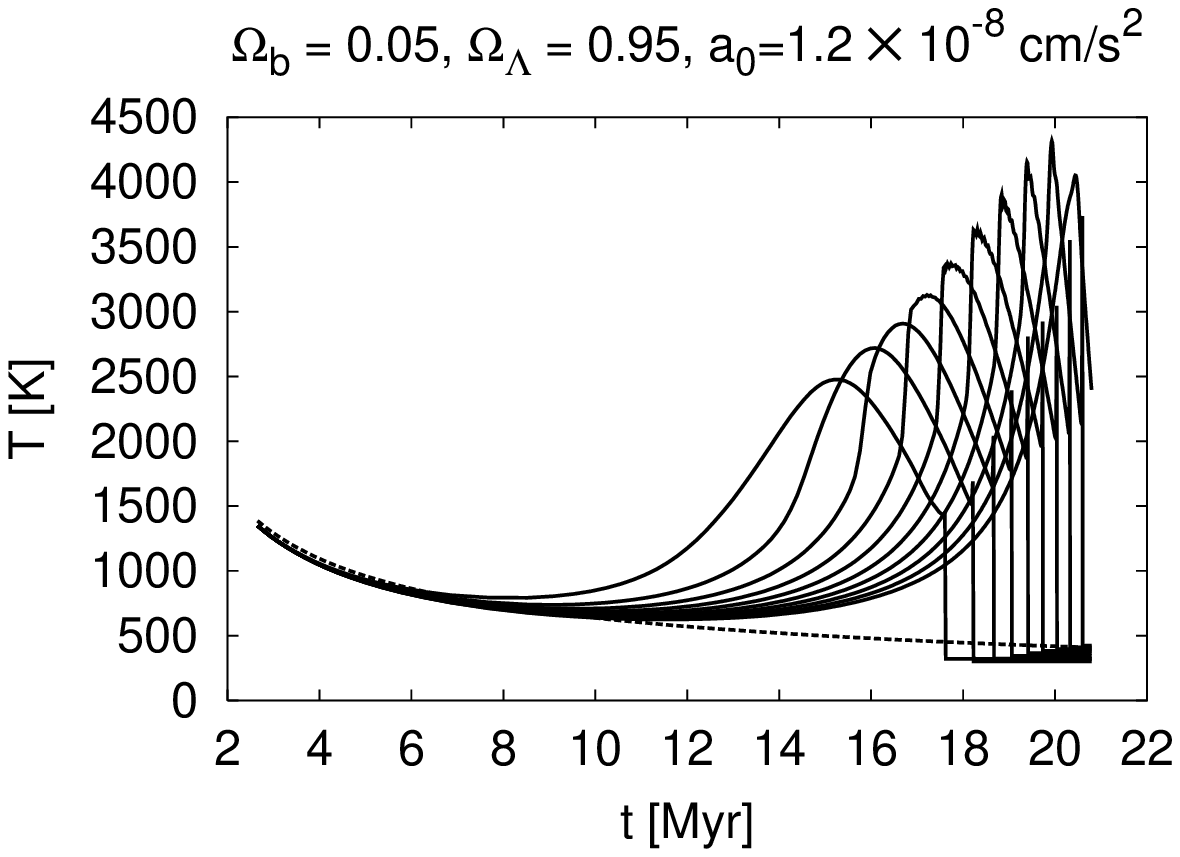}{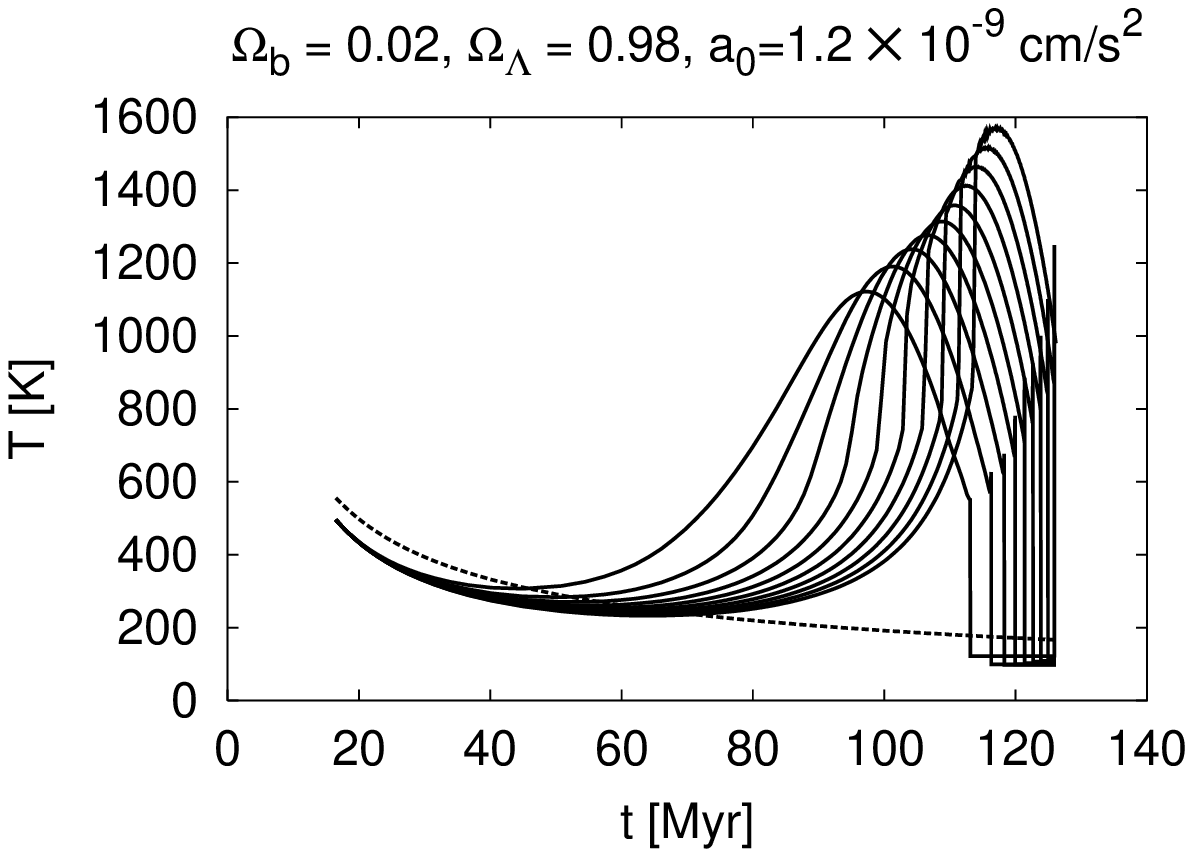}{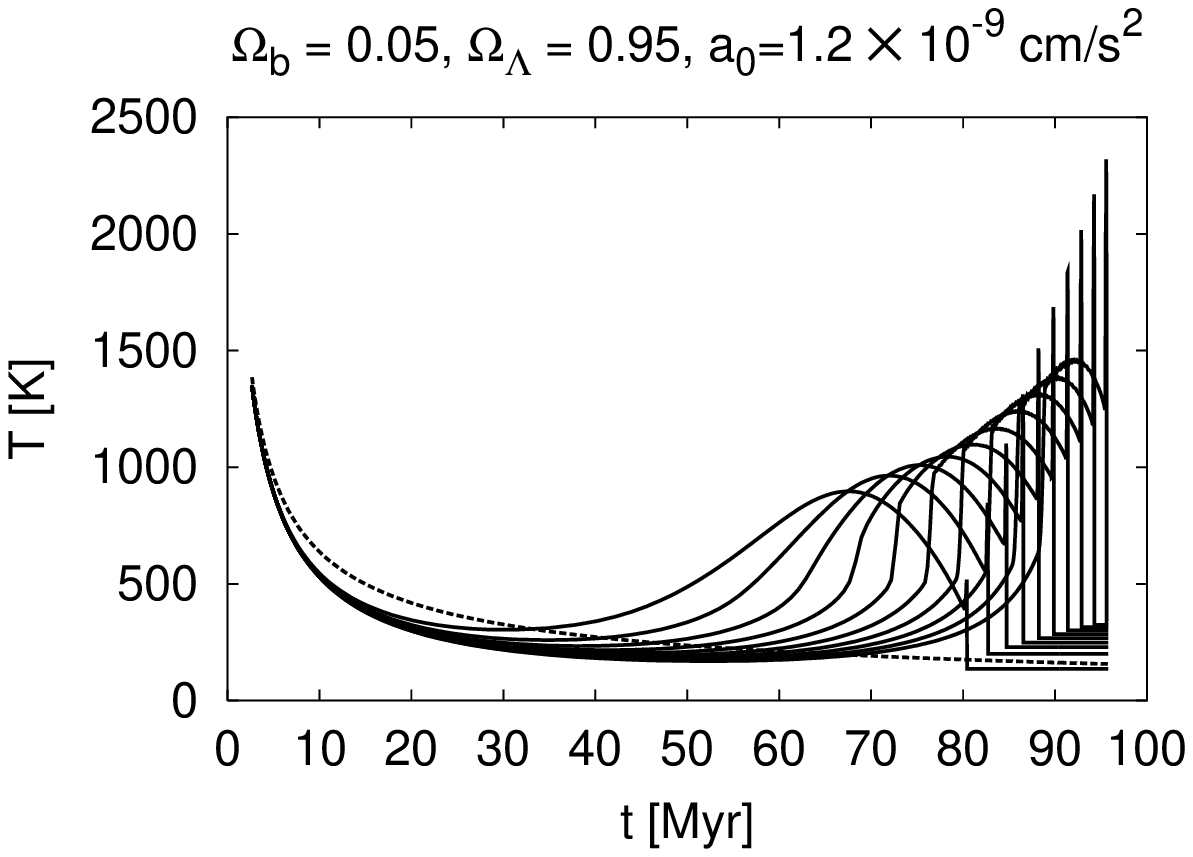}{Shell temperatures for
the runs with non-zero cosmological constant as a function of cosmic time.}

\poczwrysl{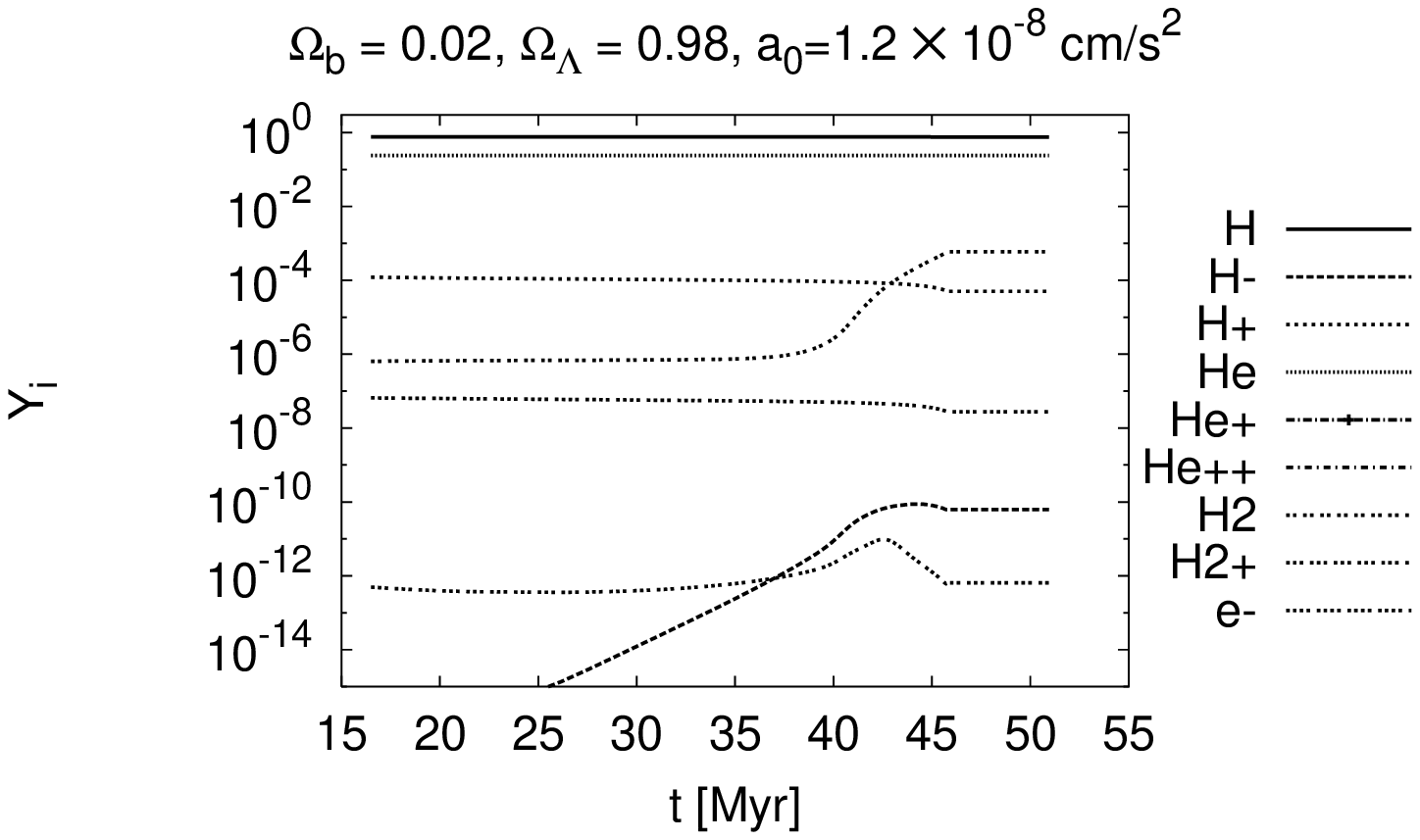}{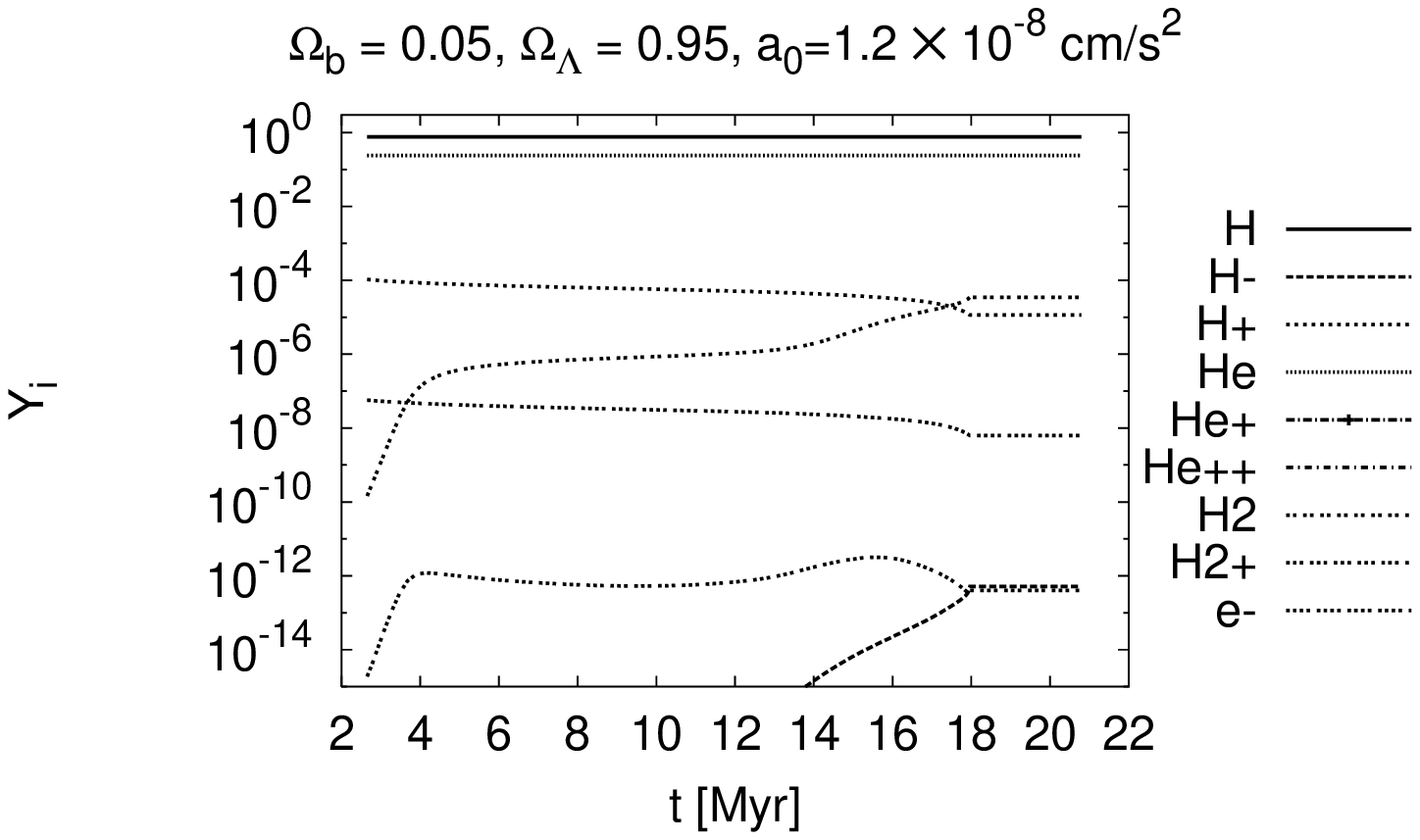}{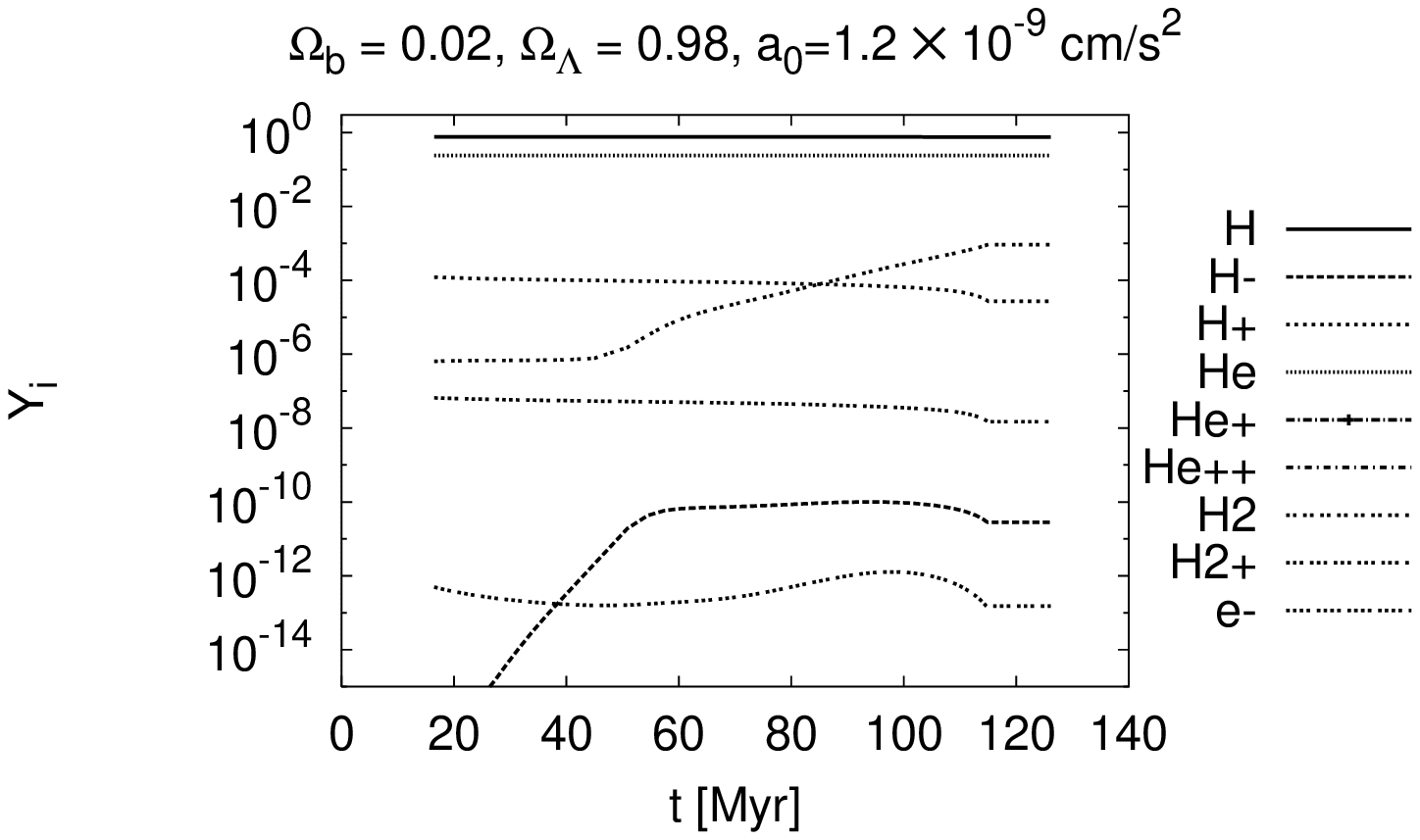}{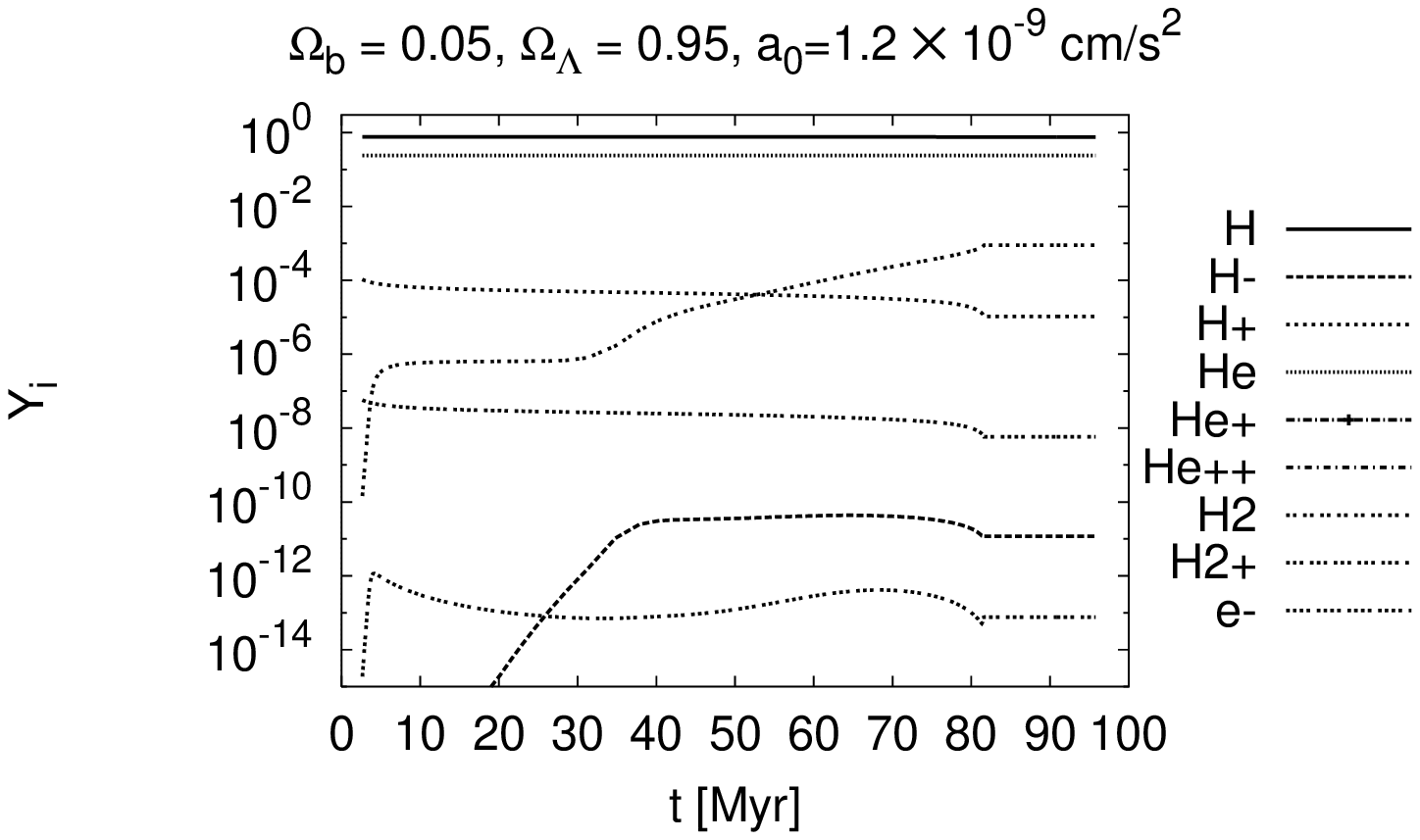}{The evolution of
chemical composition for a shell enclosing 12\% total mass, for
the runs with non-zero cosmological constant.}

The last two figures show the evolution of shell temperatures (Fig. 6) and
chemical composition of a shell enclosing 12\% of the total mass (Fig. 7)
for the runs with non-zero cosmological constant.
Upper plots are for the ``standard'' $a_0$, lower ones for $a_0/10$,
plots at left for $\Omega_b=0.02$ and plots at right for
$\Omega_b=0.05$. Shell temperatures
behave in a very similar way: they fall (due to the Hubble expansion),
then rapidly grow (collapse) and again fall (due to the cooling processes).
For the ``standard'' $a_0$ maximal temperatures are significantly higher (up to
4500 K for the high-$\Omega_b$ model) so the cooling is much more efficient
and the final collapse is faster. It is the effect of higher gravity and,
thus, higher outer pressure. The evolution of chemical composition is
qualitatively similar for all runs and
all shells. It behaves similarly to the predictions of background
cosmological models up to the time of collapse. The greatest differences
are for $H_2^+$ and especially $H_2$ which eventually reaches the final
abundance of order of $10^{-3}$. One should stress that at the time of
collapse chemical
reactions are much faster because their rates are proportional to $\varrho^2$.

It is worth to note that in order to avoid arbitrarily small timesteps, we
have adopted a numerical trick described by Thoul and Weinberg \cite{Tho95}:
if a shell falls below some radius $r_c$ it is set to some constant value and
chemical reactions and cooling are frozen. It means that the ``flat'' parts
of the curves at the end of each plot sometimes preceeded by a strange
behaviour of the shell temperature (very narrow ``peaks'' for the low-$a_0$
models) are artificial.

\section{Conclusions}

If the assumptions of the MOND were correct, the
``Dark Ages'' end very early -- about $z \sim 110-160$ for the
``standard'' $a_0$ and about $z \sim 60$ for $a_0$ 10 times lower. It may be
a good test of MOND in future because in the standard ($\Lambda$)CDM scenario
the first luminous objects may appear only for $z\sim 10-20$. However, much
more work is necessary in order to understand MOND and, in particular, the
MOND cosmology properly.

\end{document}